\documentclass[%
 reprint,
superscriptaddress,
 amsmath,amssymb,
 aps,
]{revtex4-2}

\usepackage{soul}
\usepackage{tikz}
\usetikzlibrary{shapes.geometric, arrows}
\usepackage{graphicx}

\usepackage{dcolumn}
\usepackage{bm}
\usepackage{hyperref}

\usepackage[version=4]{mhchem}
\usepackage{siunitx}
\usepackage{comment}
\usepackage{braket}
\usepackage{booktabs}
\usepackage[multiple]{footmisc}

\DeclareSIUnit\angstrom{\text {Å}}

\tikzstyle{arrow} = [thick,->,>=stealth]

\tikzstyle{rounded_box} = [rectangle, rounded corners,
minimum width=2cm, 
minimum height=1cm, 
text centered, 
text width=2cm, 
draw=black,
fill=red!30]

\tikzstyle{box} = [rectangle, 
minimum width=3.5cm, 
minimum height=1cm, 
text centered, 
text width=3.5cm, 
draw=black,
fill=blue!30]

\tikzstyle{result} = [ellipse, 
minimum width=1.5cm, 
minimum height=1cm, 
text centered, 
text width=1.5cm, 
draw=black,
fill=green!30]

\usepackage{csquotes}

\begin{document}

\title{Interplay between Electronic Structure, Chemical Bonding, and Lattice Symmetry in Bismuth Vanadate}

\author{Philip Schwinghammer}
\affiliation{
 Physics Department, TUM School of Natural Sciences, Technical University of Munich, 85748 Garching, Germany
 }
\author{Franziska S. Hegner}
\affiliation{
 Physics Department, TUM School of Natural Sciences, Technical University of Munich, 85748 Garching, Germany
 }
  \affiliation{Walter Schottky Institute, Technical University of Munich, 85748 Garching, Germany}

  \author{Frederico P. Delgado}
\affiliation{
 Physics Department, TUM School of Natural Sciences, Technical University of Munich, 85748 Garching, Germany
 }

  \author{Michel Panhans}

  \affiliation{
Department of Chemistry, TUM School of Natural Sciences, Technical University of Munich, 85748 Garching, Germany
 }

  \author{Konrad Merkel}

  \affiliation{
Department of Chemistry, TUM School of Natural Sciences, Technical University of Munich, 85748 Garching, Germany
 }

\author{Frank Ortmann}

\affiliation{
 Department of Chemistry, TUM School of Natural Sciences, Technical University of Munich, 85748 Garching, Germany
 }

\affiliation{
 Atomistic Modeling Center, Munich Data Science Institute, Technical University of Munich, 85748 Garching, Germany
 }

 \author{Ian D. Sharp}
 \affiliation{
 Physics Department, TUM School of Natural Sciences, Technical University of Munich, 85748 Garching, Germany
 }
 \affiliation{Walter Schottky Institute, Technical University of Munich, 85748 Garching, Germany}
\author{David A. Egger}
\email{david.egger@tum.de}
\affiliation{
 Physics Department, TUM School of Natural Sciences, Technical University of Munich, 85748 Garching, Germany
}
\affiliation{
 Atomistic Modeling Center, Munich Data Science Institute, Technical University of Munich, 85748 Garching, Germany
 }
\date{\today}

\begin{abstract}

Bismuth vanadate (\ce{BiVO4}) is a prototypical oxide photocatalyst that occurs in both tetragonal and monoclinic scheelite phases with markedly different photocatalytic and photoelectrochemical activities.
Accurately identifying the monoclinic phase as the ground state and explaining the origin of its symmetry-breaking distortion are unusually challenging from a theoretical perspective, with various levels of theory and associated physical interpretations for this behaviour reported in the literature. Here, we resolve these discrepancies by systematically assessing the role of exact exchange with and without spin-orbit coupling, demonstrating that an accurate treatment of electronic localization is essential to stabilize the monoclinic scheelite structure. Using this framework, we compute the electronic band structure through dense sampling of the  Brillouin zone and show that the band edges in monoclinic and tetragonal \ce{BiVO4} lie far from conventional high-symmetry paths, leading to substantial differences in band gaps and carrier effective masses. Choosing the exchange-correlation functional that best reproduces the crystal structure leads to excellent predictions of the band gap once excitonic and thermal effects are taken into account. In addition, we show that the monoclinic distortion is driven by charge transfer between non-equivalent oxygen sites, which breaks the lattice symmetry and is suppressed by self-interaction errors when using semi-local DFT. These results establish a direct connection between the exchange-correlation functional, electronic localization, chemical bonding, and structural stability in \ce{BiVO4}, providing a foundation for robust \textit{ab initio} descriptions of phase stability and optoelectronic properties in such complex oxides.

\end{abstract}

\maketitle
\thispagestyle{plain}

\section{\label{sec:level0}Introduction}

Bismuth vanadate (\ce{BiVO4}) is a promising oxide semiconductor that has been intensively investigated for solar-to-chemical energy conversion, including photocatalytic and photoelectrochemical water-splitting \cite{Park2013, Sharp2017}. Among oxide-based light absorber materials, \ce{BiVO4} exhibits strong visible light absorption and a high activity for driving various oxidation reactions \cite{Abdi2013, Qi2022, Lu2024, Zhou2025}. Despite these favorable characteristics, \ce{BiVO4} suffers from sluggish and poorly understood charge carrier transport properties. In particular, electrons are believed to form small polarons \cite{Kweon2015}.
The nature of holes, however, remains incompletely understood \cite{Kweon2013, Kweon2013b, Liu2020}.
Therefore, improving the performance of this prototypical photocatalytic material requires a detailed understanding of its exact electronic and lattice structure, as well as how their interplay impacts the functional properties. 

Relevant in this context, \ce{BiVO4} can be found in several different crystal structures, including the monoclinic scheelite, tetragonal scheelite, and tetragonal zircon polymorphs \cite{NAGABHUSHANA2015}. Among these, the monoclinic scheelite phase is the most promising for photochemical energy conversion \cite{Kudo99}, and has been identified as the room temperature ground state structure \cite{SLEIGHT1979}. Upon heating to 528~K, \ce{BiVO4} undergoes a second order displacive phase transition to the tetragonal scheelite structure \cite{SLEIGHT1979}. The tetragonal scheelite phase, though physically similar to the monoclinic structure, exhibits significantly poorer photocatalytic activity, though the underlying origin of this difference is not yet completely resolved.\cite{Tokunaga2001, Iwase2010}.
In the following, references to monoclinic and tetragonal \ce{BiVO4} refer specifically to the scheelite polymorphs.

Answering the question of why monoclinic \ce{BiVO4} exhibits superior performance for solar-to-chemical energy conversion requires a detailed theory of the interplay between electronic structure, chemical bonding, and lattice symmetry. However, a comprehensive framework connecting these effects remains elusive. It has been suggested that symmetry breaking in the monoclinic phase leads to a hybridization of empty Bi6p states with occupied (Bi\,6s--O\,2p)$^*$ at the valence band edge, which reduces the total energy and stabilizes monoclinic \ce{BiVO4} relative to the tetragonal phase, as first noted in ref. \cite{Walsh2009} and later explained in detail in ref. \cite{Kweon2012}. However, this explanation neglects the impact of ionic bonding and V--O bonds, as well as the presence of occupied Bi\,6p--O\,2p states, and therefore remains incomplete. 

These issues are further exacerbated by the limitations of common theoretical approaches for accurately describing both the electronic and lattice structures of \ce{BiVO4}, which remain contentious. A key illustration of this difficulty is the work of Kweon \textit{et al.}, who showed that relaxation to the ground state structure with semilocal density functional theory (DFT) mistakenly predicts tetragonal scheelite as the lowest-energy structure of \ce{BiVO4} \cite{Kweon2012}. One solution to overcome this issue is the use of hybrid DFT functionals, which mix semilocal and Fock-exchange \cite{Kweon2012, Liu2022}. Two common variants include global hybrids, such as PBE0 with an exact-exchange fraction of $\alpha = 0.25$ \cite{Perdew96b,Ernzerhof99,Adamo99}, and range-separated hybrids such as HSE06 \cite{Heyd2003, Krukau06}, which retains the same exact exchange fraction but restricts it to distances of approximately $\leq \SI{4.8}{\angstrom}$. Kweon \textit{et al.} found that use of HSE06 provides good agreement with experimental data for the ground state structure \cite{Kweon2012}. While later work by Liu \textit{et al.} confirmed that semilocal DFT yields an incorrect ground state structure, they were not able to reproduce the success of the HSE06 functional, instead settling on a global hybrid with an unusually large value of $\alpha = 0.6$ \cite{Liu2022}. Ohad \textit{et al.} investigated the use of Wannier-localized optimally-tuned screened range separated hybrids (WOT-SRSH) in ref. \cite{Ohad2023}, using a short range exact exchange fraction of $\alpha = 0.25$ as an ansatz, and tuning the long-range fraction to $\beta = \epsilon_{\infty}^{-1} = 0.17$, but did not investigate the impact on the structure. 
 
Over the years, a wide range of different functionals have been used to account for self-interaction errors in the description of electronic, structural, and polaronic properties of \ce{BiVO4} \cite{Kweon2012, Wadnerkar2013, Kweon2013, Kweon2015, Wiktor2017, Pasumarthi2019, Liu2020, Liu2022, Ohad2023}. Although several explanations have been proposed for why the exact-exchange fraction influences the predicted ground state structure of \ce{BiVO4}\cite{Kweon2012, Liu2022}, a comprehensive understanding of the underlying physical origin of this behavior remains lacking. Liu \textit{et al.} noted that spin-orbit coupling (SOC) increases the energy difference between monoclinic and tetragonal structures at high levels of exact exchange \cite{Liu2022}, but did not investigate its impact on symmetry breaking and the ground state structure. In sum, neither the appropriate level of theory for accurately predicting the lattice and electronic structure of \ce{BiVO4}, nor the role of exact exchange and SOC on bonding and monoclinic phase stabilization, has been established. In order to provide the tools for a foundational theoretical description of the material, we explain the fundamental bonding mechanisms and how they relate to the approximations made to describe the electrons, culminating in a reliable and physically justified computational approach.

These methodological issues are directly relevant for understanding the optoelectronic functionality of \ce{BiVO4} because they strongly influence the outcomes of first-principles calculations used to interpret experiments. Indeed, even the theoretical literature lacks consensus on several basic aspects of the electronic band structure. 
One important example is the nature of the fundamental band gap, one of the most important properties of any semiconductor. In the case of \ce{BiVO4}, an early theoretical study predicted a direct band gap \cite{Walsh2009}, consistent with experimental interpretations at the time \cite{Stoughton2013}. However, subsequent theoretical and experimental studies provided compelling evidence for an indirect electronic band gap \cite{Ding2013, Wang2014, Cooper2015}, indicating that precise $k$-sampling is required for analysis of the extended band structure of this material. Furthermore, the use of hybrid functionals \cite{Wadnerkar2013}, as well as physical effects such as SOC and finite temperature \cite{Wiktor2017}, may influence the numerical value of the band gap. Finally, in order to make accurate predictions of the absorption spectrum, excitonic effects must be taken into account, in order to predict not only the fundamental, but also the optical band gap. 

Worse uncertainty persists regarding the dispersion near the band edges. While previous work reported DFT-calculated effective masses \cite{Zhao2010, Zhao2011, Ding2013, Lardhi2018,Ghule2026}, these studies used semilocal DFT, relaxing \ce{BiVO4} to a quasi-tetragonal symmetry, which is not the true $\SI{0}{K}$ structure~\cite{Kweon2012, Liu2022}. 
Furthermore, previous ab initio estimates of effective masses in \ce{BiVO4} were obtained via fits along high-symmetry directions and did not account for SOC \cite{Walsh2009, Zhao2010, Zhao2011, Ding2013, Lardhi2018, Newhouse2018, Ghule2026}. These issues are important because the dispersion at the electronic band edges strongly influences carrier transport, both via the direct dependence of band-like mobility and through charge localization and polaron formation. 
For example, poor electron transport in \ce{BiVO4} has been attributed to small electron polaron formation \cite{Kweon2015}, or to large effective masses and defect formation \cite{Lardhi2018}. In addition, there remain conflicting predictions regarding the degree of localization of holes as large or small polarons \cite{Kweon2013, Kweon2013b, Wiktor2018, Pasumarthi2019, Liu2019, Liu2020, Hao2025}. Thus, determination of the band edge dispersion and effective masses provides important insights into charge carrier localization, polaron formation, and the differing functional properties of monoclinic and tetragonal \ce{BiVO4}, all of which are relevant for understanding energy conversion processes in these materials.

In this work, we systematically investigate the role of hybrid functionals and SOC in determining the structural and optoelectronic properties of \ce{BiVO4} and clarify the microscopic mechanism stabilizing the monoclinic phase. 
We provide a detailed account of the electronic band structures of both monoclinic and tetragonal phases, including a precise determination of the locations of the band edges in the Brillouin zone, the band gaps, and carrier effective masses.
Through the calculation of absorption spectra, we assess the impact of symmetry breaking on the optical gap and find significant differences between the fundamental band structures of monoclinic and tetragonal \ce{BiVO4} but only small differences for the excitonic effects.
We find that charge transfer between non-equivalent oxygen sites acts as the central mechanism for symmetry breaking of \ce{BiVO4}, and examine how this behavior is impacted by the fraction of exact exchange used in the DFT functional. Together with analysis of the impact of SOC on the predicted crystal structure of \ce{BiVO4}, we provide a rigorous description of the interplay between electronic structure, chemical bonding, and lattice symmetry in this important photocatalytic material. 

\section{\label{sec:level1}Methods}

We performed DFT calculations using VASP \cite{Vasp1,Vasp2,Vasp3}, employing the projector-augmented-wave method to describe the electron–ion interaction \cite{Blöchl1994}, with valence configurations of 5d$^{10}$6s$^2$6p$^3$ for Bi, 3p$^6$3d$^3$4s$^2$ for V, and 2s$^2$2p$^4$ for O. The plane-wave cutoff was converged to $550$ eV, with a $4 \times 4 \times 2$ $k$-grid for the monoclinic and tetragonal conventional cells, and a $5 \times 5 \times 7$ $k$-grid for the monoclinic and tetragonal primitive cells, sampling the $k$-points with the Monkhorst-Pack scheme \cite{Monkhorst1976}. We used Gaussian smearing with a width of $0.03$ eV and the Blocked-Davidson algorithm for all calculations with semilocal functionals. For the hybrid DFT calculations, this setup was combined with the Adaptively Compressed Exchange (ACE) operator \cite{Lin2016}, unless numerical convergence was difficult, in which case we used the damped velocity friction algorithm with a time step of 0.3.

\subsection{\label{sec:level1a}Geometry optimization}

All geometry optimizations were performed using the monoclinic and tetragonal conventional cells for ease of comparison. We performed relaxations using a global hybrid functional equivalent to PBE0 but varied the exact-exchange fraction, $\alpha$, between 0.0 and 0.6 using a step size of 0.1, and repeated the same procedure for HSE06, additionally including the canonical value of $\alpha=0.25$ for both. We performed such calculations while also accounting for SOC using selected Fock-exchange fractions of $\alpha = 0.0, 0.25, 0.4$, and $0.6$. In all cases, the total energy was converged to $10^{-7}$~eV, $k$-space sampling was not reduced due to symmetry, and all atomic and lattice degrees of freedom were optimized. The geometry relaxations were performed with the GADGET code from ref. \cite{Bucko05}, using a convergence criterion of $10^{-4}$~Hartree/bohr for the forces. We used the BFGS update formula for the Hessian and the GDIIS algorithm for optimization. The covalent radii for identification of short-range and long-range internal coordinates were scaled by factors of $1.2$ and $2.2$, respectively. 

\subsection{\label{sec:level1c}Electronic band-structure calculations}

In order to identify the correct valence band maximum (VBM) and conduction band minimum (CBM) of monoclinic and tetragonal \ce{BiVO4}, we first performed regular DFT calculations using the HSE06+SOC approach for the primitive cells, applying a standard $5 \times 5 \times 7$ Monkhorst-Pack $k$-mesh. These calculations were followed by self-consistent calculations with zero-weighted high symmetry lines, generated using VASPKIT \cite{VASPKIT}, added to the regular k-mesh, restarted from the wave functions of the previous DFT run. 
We identified the approximate VBM and CBM locations from these calculations and subsequently performed a grid search across the entire Brillouin zone. Here, the band structure was self-consistently calculated in a cube surrounding the VBM and CBM identified in the previous step, sampled by a $3 \times 3 \times 3$ $k$-grid of zero-weighted points added to the regular k-mesh, and the procedure was repeated until convergence. These calculations were performed for successively finer grids and correspondingly smaller cubes, beginning with a grid spacing of 0.01 in fractional coordinates and culminating in a calculation with grid spacing of 0.002 in fractional coordinates. 
We then calculated the effective mass tensor for the resulting VBM and CBM using the effective mass calculator from ref. \cite{afonari}. This code uses a finite-difference approach to calculate the second derivative of the band energies with respect to $\Vec{k}$. For the hybrid DFT calculations, we slightly modified the workflow by treating the additional $k$-points for the finite difference calculations self-consistently with zero-weighted $k$-points, analogous to band-structure calculations with hybrid functionals. For the additional $k$-grid used for the finite difference calculations, we used the 5-point variant provided by the code. 
For the calculation of the energy isosurfaces in k-space, we used the band structure from the previously mentioned high-symmetry path calculations, and post-processed it using the IFermi code \cite{Ganose2021}. The markers for the VBM and CBM were placed using the locations previously calculated, and applying all relevant symmetry operations for each structure. 

\subsection{\label{sec:level1d}Calculation of excitonic and thermal corrections to the band gap}

To evaluate the optical properties of \ce{BiVO4}, including excitonic effects, we computed the optical response within the Bethe--Salpeter equation (BSE) using the linear-scaling Wannier-Optics code~\cite{Merkel_2024,WannierOptics_code}. In this approach, the BSE is formulated in terms of an effective electron--hole Hamiltonian represented in a localized Wannier basis. The method corresponds to the BSE in the Tamm--Dancoff approximation (TDA), but avoids an explicit representation of the full excitonic Hamiltonian in the conventional reciprocal-space transition basis. Instead, the relevant Hamiltonian matrix elements are evaluated in a basis of maximally localized Wannier functions (MLWFs)~\cite{Marzari1997,Marzari2012}.

To obtain the MLWFs, we first calculated the electronic structure using VASP on a $\Gamma$-centered $5\times5\times3$ \textit{k}-point grid, with a plane-wave energy cut-off of 550 eV and the PBE exchange--correlation functional. The MLWFs were then generated with the \textsc{wannier90} code \cite{Pizzi2020} using the VASP2Wannier interface. In the present case, 48 valence and 20 conduction-band Wannier functions were obtained by separately Wannierizing the occupied and unoccupied subspaces. Further details of the linear-scaling Wannier-Optics method are given in Ref.~\cite{Merkel_2024}.

For calculation of the absorption spectra of \ce{BiVO4}, we considered the diagonal components of the macroscopic, frequency-dependent dielectric tensor in the long-wavelength limit $ \epsilon_{jj}(\omega) $. Within the linear-scaling Wannier-Optics approach, the optical response is evaluated by propagating the electron--hole polarization in time and subsequently transforming the resulting response to the frequency domain. Specifically, the dielectric function was calculated as
\begin{equation}
    \epsilon_{jj}(\omega) = 1 - \frac{8\pi}{\Omega \hbar}\int_0^{\infty}dt e^{i(\omega + i\eta)t}\Im(\braket{\Psi^{(j)}(0)|\Psi^{(j)}(t)}), \label{eq:dielectric function}
\end{equation}
where the initial state $\ket{\Psi^{(j)}(0)}$ is generated by the Cartesian $j$-component of the dipole-operator acting on the many-body ground state, $\Omega$ is the volume of the unit cell, and $\eta$ is a small positive broadening parameter. 
This time-propagation scheme provides a numerically efficient route to the TDA-BSE absorption spectrum while avoiding an explicit diagonalization of the excitonic Hamiltonian.

For the static screening of the Coulomb interaction in \ce{BiVO4}, we used a dielectric constant of 5.35. This value is the spatial average of the principal values of the static dielectric tensor calculated via finite differences, as described in ref. \cite{Souza2002}. For consistency, the static tensor was evaluated using HSE06+SOC for the primitive monoclinic ground-state structure, which had been relaxed at the same HSE06+SOC level. For these calculations, we used a $5\times5\times7$ $k$-grid with otherwise identical numerical settings, and a finite electric field of $\SI{0.001}{eV/\angstrom}$. The value for the dielectric constant is in decent agreement with the value of 5.92 from \cite{Ohad2023}, where they used a WOT-SRSH functional. In order to align the dielectric function calculated using PBE-based band structures with the results obtained using hybrid DFT, we applied a scissor shift equal to the difference between the fundamental gap calculated with HSE06+SOC and calculated with PBE. 

In order to account for changes to the band gap due to zero-point quantum fluctuations and finite temperature lattice effects, we performed Monte Carlo simulations using the one-shot method from ref. \cite{ZG_2016} with a nondiagonal monoclinic supercell containing eight primitive cells, corresponding to a $2 \times 2 \times 1$ supercell of the conventional cell, on a $2 \times 2 \times 2$ $k$-grid. Due to the relatively large size of the supercell (96 atoms), the calculations were considered sufficiently converged, as calculations for larger cells are not feasible with HSE06+SOC. The band structure was evaluated for each distorted supercell, as well as for the original pristine supercell, using the same $k$-grid. Since the tetragonal structure is not dynamically stable, Monte-Carlo sampling cannot be used for tetragonal \ce{BiVO4} directly. Therefore, we assume the temperature-induced band-gap correction to be the same as that of the monoclinic structure, justified by the nearly identical structures. 

\subsection{\label{sec:level1b}Bonding Analysis}

For the bonding analysis, we primarily used the LOBSTER code \cite{Lobster1, Lobster2, Lobster3}. Since LOBSTER is not compatible with SOC, we used HSE06 with $\alpha = 0.4$, since that Fock-exchange fraction best reproduces the experimental structure without SOC. The electronic structure was recalculated with a larger number of bands, converging the  total energy to a precision of $10^{-6}~$eV.  
With the LOBSTER code, we transformed the DFT-Hamiltonian into a local atomic basis and calculated the projected crystal orbital Hamilton population (pCOHP) among the different Bi and O orbitals. The pCOHP provides an energy-resolved measure of the strength of bonding and antibonding interactions between specific atomic orbitals. It should be noted that the choice of orbitals, and in particular the choice of whether to orthonormalize the atomic basis, may have an impact on the resulting pCOHP \cite{oliphant2026}. Since our analysis relies partially on quantifying the relative contributions from different atomic orbitals to the eigenstates, using a non-orthogonal basis would lead to ill-defined results. Nevertheless, the choice of atomic basis functions may alter the relative importance of covalent and ionic bonding in a material. We therefore only analyze the qualitative impact of each on the phase transition within a single framework. 

\begin{figure*}[htb]
\centering
\includegraphics[trim={1.4cm 4cm 1.4cm 2cm},clip,width = \textwidth]{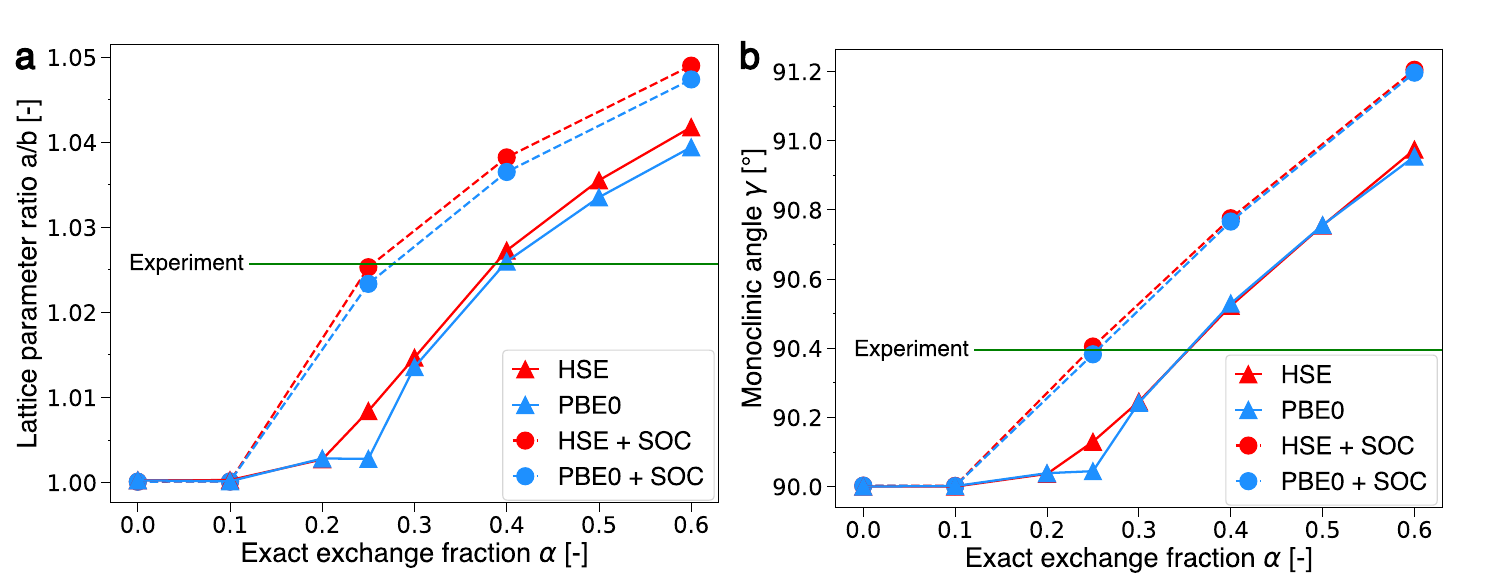}
\caption{(a) DFT-calculated ratio of lattice constants $a$ and $b$ and (b) monoclinic angle $\gamma$ as a function of the exact exchange fraction $\alpha$ used in the HSE06 and PBE0 functional, with and without spin-orbit coupling (SOC). The horizontal green line in each panel indicates experimental values for monoclinic \ce{BiVO4} reported in ref. \cite{SLEIGHT1979}. Tetragonal \ce{BiVO4} corresponds to a ratio of $a$ and $b$ of 1.0 and to $\gamma=90°$, respectively.}
\label{fig:Monoclinic_Angle}
\end{figure*}

In order to create a simplified bonding diagram, we used the LOBSTER code to transform the DFT-Hamiltonian into an atomic basis, yielding a tight-binding (TB) Hamiltonian using the same procedure employed for the pCOHP calculations described above. Following this, we first truncated the TB Hamiltonian to include only the states with significant contributions near the VBM and CBM, and diagonalized the resulting Hamiltonian in the reduced basis to obtain the eigenstates. To obtain a representation more closely aligned with chemical intuition, we then grouped the manifold of eigenenergies across all $k$-points according to both their energy and the dominant atomic orbital character of each eigenstate, resulting in a representation with far less energy levels. The first step was calculating the DOS (with gaussian smearing using a width of \SI{0.3}{eV}), and separating the energy into subranges by making cuts at each minimum in the DOS. Each atomic orbital $\ket{\phi_i}$ was assigned to one physically significant type $t$ of orbital, including $t=\{$ Bi6p, Bi6s, V3d, O1-2p, and O2-2p$\}$, where O1 and O2 denote the two oxygen sites which are not symmetry equivalent in the monoclinic phase. We then calculated orbital weights for each type, $t$, and eigenstate, $\ket{\Psi_n (\Vec{k})}$, according to $w_{n,t}(\Vec{k}) = \sum_{i \in t} |\braket{\phi_i|\Psi_n(\Vec{k})}|^2$. Inside each energy subrange, the eigenstates where then grouped by type using a complete-linkage clustering algorithm \cite{Müllner2011} implemented in scipy \cite{Virtanen2020}, with the distance function
\begin{equation}
    d(\Vec{w}_n(\Vec{k}), \Vec{w}_m(\Vec{k}')) = \frac{1}{\sqrt{2}}\sqrt{\sum_t \left(w^{\gamma}_{n,t}(\Vec{k}) - w^{\gamma}_{m,t}(\Vec{k}')\right)^2},
\end{equation}
where the exponent $\gamma$ is a parameter tuned to emphasize small contributions, and the clustering algorithm is controlled by the cutoff parameter $d_{cut}$. In the present work, the parameter $\gamma = 0.1$ and $d_{cut} = 0.3$ were chosen to minimize the number of visible effective states without oversimplifying. 

Within each group, we averaged over all levels separated by no more than $\SI{1}{eV}$, and set weights lower than 0.2$\%$ to zero. To construct the bonding diagram, we then used the average on-site energies of each orbital type to represent the atomic orbital energies, and the simplified TB energy levels to represent the molecular orbital energies. Each molecular orbital was then plotted in colored segments corresponding to the normalized square root of the weights $w_{n,t}$ contributing to that orbital. 

Finally, in order to analyze the localization of charge in the monoclinic and tetragonal structures, we used the Bader-charge analysis code by the Henkelmann group \cite{Henkelman2006, Tang2009, Sanville2007, Yu2011}, utilizing the same DFT settings as for the LOBSTER calculations.

\section{\label{sec:level2}Results}

In order to understand the interplay between lattice symmetry and electronic structure in \ce{BiVO4}, we first identify an appropriate exchange-correlation functional by comparing DFT-computed geometries to experimental data. We then evaluate the impact of the monoclinic distortion on the electronic structure, demonstrating how changes in the crystal symmetry modify the band gap, effective masses, band edge degeneracies, and excitonic effects. Finally, we analyze the chemical bonding responsible for the structural distortion, showing how symmetry breaking enables both covalent and ionic energy-lowering mechanisms.

\subsection{\label{sec:level2a}Impact of exact exchange on geometry optimization}

To compare the DFT-predicted ground-state structures with experiment, we first evaluated the lattice constant ratio, $\frac{a}{b}$, and the monoclinic angle, $\gamma$, for different exchange-correlation functionals and exact-exchange fractions. In the tetragonal structure of \ce{BiVO4}, these quantities are equal to 1.0 and $90$°, respectively. $\frac{a}{b}$ and $\gamma$ increase monotonically with increasing monoclinic distortion, and they are invariant under rotations and translations of the conventional cell as well as isotropic scaling of the compared structures. Hence, they provide a robust measure for how accurately the experimentally observed symmetry-breaking is reproduced by the calculations. 

We find that both the fraction of exact exchange and the inclusion of SOC have profound impacts on the DFT-computed crystal structure. 
Specifically, use of the semilocal PBE functional, corresponding to $\alpha=0.0$, leads to relaxation into the tetragonal structure (see Figure  \ref{fig:Monoclinic_Angle}), consistent with previous reports \cite{Kweon2012, Liu2022} and irrespective of whether SOC is included. Without including SOC, increasing the exact-exchange fraction leads to a gradual increase of the structural parameters associated with the monoclinic distortion. However, use of $\alpha = 0.25$ fails to reproduce the experimental values, in agreement with ref. \cite{Liu2022}, which likewise was unable to reproduce the findings of ref. \cite{Kweon2012}. Ultimately, we find the best agreement with experimental values of $\frac{a}{b}$ and $\gamma$ using $\alpha \approx 0.35$ and $\alpha \approx 0.4$, respectively. The inclusion of range-separation had no significant effect on the calculated structural parameters associated with the degree of the monoclinic distortion, apart from a small outlier at $\alpha = 0.25$. 

Our estimate for the ideal value of $\alpha \approx 0.4$ is significantly lower than that of $\alpha = 0.6$ which Liu \textit{et al.} suggested in ref. \cite{Liu2022}. This is partially due to a discrepancy in the DFT-calculations: evaluating the mean-square deviation (MSD) of the Bi--O bond lengths to quantify agreement with experiment the same as they did, we find a best agreement of $\alpha = 0.5$ when using PBE0. Overall, it appears their calculations increase the required $\alpha$ by about 0.1 when compared to ours purely based on the difference in DFT-results. However, part of the discrepancy is also because Liu \textit{et al.} did not compare the agreement of the V--O bond lengths with experiment. When we calculate the full MSD of all bond lengths, we consistently find an ideal value of $\alpha \approx 0.4$ for PBE0- and HSE06-based calculations. This agreement with the lattice-parameter-based metric implies that the most physically sound and consistent comparison to experiment leads to a lower value of $\alpha = 0.4$. While it does not fully resolve the discrepancy between our results and those of Liu \textit{et al.}, it accounts for a large part of the offset, and partially explains why their optimal value for $\alpha$ is larger than ours. A more in-depth comparison of the DFT-calculations and convergence tests can be found in the SI.

Nevertheless, since the experimental parameters for $\frac{a}{b}$ and $\gamma$ are reproduced for different fractions of exact exchange, there is no single value for $\alpha$ which perfectly describes the overall geometry of \ce{BiVO4}. This inability to reproduce experiment with a single value for $\alpha$ is resolved once SOC is included. We find that SOC increases the monoclinic distortion observed at all values of $\alpha$, leading to nearly perfect agreement between the DFT-computed and experimentally-determined structural observables for quantifying the distortion ($\frac{a}{b}$ and $\gamma$), at a value of $\alpha = 0.25$. Likewise, both the Bi--O bond length deviation and the averaged deviation over all bonds is minimized compared to experiment for $\alpha = 0.25$ when SOC is taken into account. While the agreement of the V--O bond lengths with experiment becomes slightly worse compared to those obtained with $\alpha = 0.0$, the actual mean square deviation remains extremely small at only $\SI{1.8E-3}{\angstrom^2}$, which represents excellent overall agreement. In addition, inclusion of SOC significantly improves the agreement between the experimental and theoretical values of the crystalline unit cell volume. In particular, without SOC, the calculated volume is underestimated by 1.6\%, whereas including SOC with $\alpha = 0.25$ leads to an underestimation by only $0.4\%$ compared to the monoclinic structure measured at 4.5~K. 

While we find that agreement with experiment is slightly improved when using HSE06, the differences from results obtained using PBE0 remain small. In particular, when SOC is included, the deviations of the lattice constant ratio, monoclinic angle, and average bond length MSD between experiment and HSE06 are 0.034\%, 0.012\%, and $\SI{0.8E-3}{\angstrom^2}$, respectively. In contrast, PBE0 yields deviations of 0.225\%, 0.013\%, and \SI{1.1E-3}{\angstrom^2}, respectively. The computed lattice parameters using the optimized exact-exchange fraction with HSE06+SOC are presented in Table \ref{tab:structure_params}.
The changes to all Bi--O and V--O bonds induced by the phase transition are explained later in the text and shown in Fig. \ref{fig:crystal_changes}.
We conclude that using a hybrid functional with $\alpha \approx 0.25$ and including SOC allows for an accurate description of the potential energy surface and geometry of \ce{BiVO4}.

\begin{table}
\centering
\caption{Summary of the structural parameters for monoclinic and tetragonal \ce{BiVO4} calculated using HSE06+SOC with $\alpha = 0.25$.}
\label{tab:structure_params}
\vspace{2mm}
\begin{tabular}{c|c|c}
    \toprule
        & \ce{monoclinic} & \ce{tetragonal} \\
    \midrule
        $a$ & \SI{5.20}{\angstrom} & \SI{5.12}{\angstrom} \\
        $b$ & \SI{5.07}{\angstrom} & \SI{5.12}{\angstrom} \\
        $c$ & \SI{11.73}{\angstrom} & \SI{11.64}{\angstrom} \\
        $\gamma$ & \SI{90.40}{\degree} & \SI{90.00}{\degree} \\
    \midrule
        Bi--O1 (short) & \SI{2.34}{\angstrom} & \SI{2.43}{\angstrom} \\
        Bi--O1 (long)  & \SI{2.37}{\angstrom} & \SI{2.48}{\angstrom} \\
        Bi--O2 (short) & \SI{2.51}{\angstrom} & \SI{2.43}{\angstrom} \\
        Bi--O2 (long)  & \SI{2.66}{\angstrom} & \SI{2.48}{\angstrom} \\
    \midrule
        V--O1 & \SI{1.76}{\angstrom} & \SI{1.72}{\angstrom} \\
        V--O2 & \SI{1.68}{\angstrom} & \SI{1.72}{\angstrom} \\
    \bottomrule
\end{tabular}
\end{table}

\subsection{\label{sec:level2c}Impact of symmetry breaking on the band-edge dispersion}

Having determined that a hybrid functional with \mbox{$\alpha=0.25$}, combined with SOC, can accurately describe the crystal structure of \ce{BiVO4}, we next investigate the associated electronic structure in detail. Building on the excellent agreement between the calculated and experimental crystal lattices, we seek to understand the impact of symmetry breaking on the electronic band structure, focusing in particular on the relative positions of the VBM and CBM, as well as their respective full effective mass tensors. For this purpose, we employ the HSE06 functional since it provides marginally better agreement with experiment and is less computationally demanding than PBE0. Furthermore, we compare the resulting fundamental and optical band gaps calculated with these methods to those found experimentally.

\begin{figure}
\centering
\includegraphics[width = 0.9\linewidth]{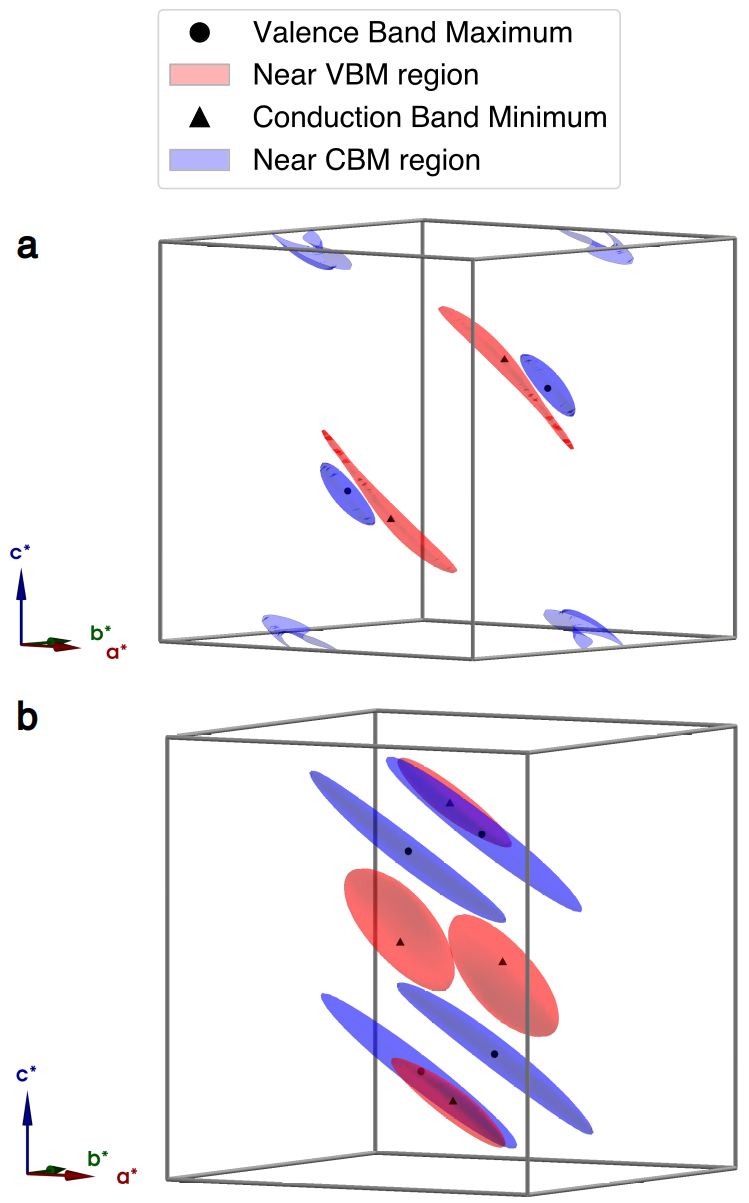}
\vspace{1ex}
\caption{Comparison of the VBM and CBM positions shown in fractional coordinates in reciprocal space, together with the thermally accessible regions near the band extrema (Constant energy surfaces at $\pm k_BT$ relative to each extremum for $T = 300$ K) for (a) the monoclinic and (b) tetragonal structures of \ce{BiVO4}. The reciprocal lattice vectors are labeled with $\Vec{a}^*, \Vec{b}^*, \Vec{c}^*$, with $\Vec{v} \cdot \Vec{u}^* = 2 \pi \delta_{v,u}$, for $u,v \in \{a,b,c\}$.}
\label{fig:band_edges}
\end{figure}

By analyzing the full Brillouin zone, we find that the VBM and CBM in monoclinic \ce{BiVO4} lie significantly closer together in $k$-space compared to their counterparts in the tetragonal structure, as given in Table \ref{tab:band_structure} and shown in Figure \ref{fig:band_edges}. In the monoclinic structure, the separation between the VBM and CBM is only $\SI{0.13}{\angstrom^{-1}}$, compared to $\SI{0.38}{\angstrom^{-1}}$ in the tetragonal phase.  This feature of the electronic structure substantially reduces the momentum transfer required for excitations across the fundamental band gap in monoclinic \ce{BiVO4} when compared to the tetragonal phase.

Having determined the positions of the band-extrema, we evaluate the dispersion of the electronic bands at these points in order to lay the foundation for a description of carrier transport and localization in \ce{BiVO4}. To characterize the band dispersion, we calculated the full effective mass tensors at the previously identified VBM and CBM positions, and list their eigenvalues in Table \ref{tab:band_structure}. We find that the effective masses are highly anisotropic in all cases, with one heavy ($\gg1m_e$), one intermediate ($\approx 1m_e$), and one light ($\ll1m_e$) mass for both electrons and holes in both the monoclinic and tetragonal phases. These results indicate that evaluating effective masses only along high symmetry directions -- as is common practice -- does not provide a good representation of the true band edge dispersion: depending on the chosen direction, \ce{BiVO4} may appear to have either exceptionally light effective masses or relatively heavy effective masses, seriously changing the expected charge carrier properties. In addition, we find that nearly all effective masses are larger in the tetragonal phase than in the monoclinic phase, with the exception of the heaviest hole mass, which is lighter. 

In addition to the band-edge dispersion, the number of symmetry-equivalent band extrema is an important quantity because the Drude model for carrier mobility based on effective masses requires averaging over all equivalent valleys.
In the bandstructure of monoclinic \ce{BiVO4},
the band extrema are mapped onto themselves by the (C$_2$) symmetry operations of the space group, which leaves the effective mass tensor unchanged. Inversion symmetry results in two distinct VBM and CBM positions in the Brillouin zone, but does not alter the effective mass tensor. 
In contrast, the tetragonal structure possesses an additional C$_4$ symmetry. In this case, the additional symmetry operations lead to four symmetry-equivalent VBMs and CBMs. Calculation of the average effective masses leaves the mass component aligned with the $\Vec{c}$ direction of the conventional cell unchanged, but increases the lowest effective masses for both holes and electrons (see Table \ref{tab:band_structure}). Accounting for this effect further enhances the differences between the two phases of \ce{BiVO4}. In particular, the lowest values for both the electron and hole effective masses in tetragonal \ce{BiVO4} are approximately twice as large as those in monoclinic \ce{BiVO4} (see Table \ref{tab:band_structure}), suggesting a greater bottleneck for band-like transport in the tetragonal phase. 

The effective masses are central to prediction of mobilities in the Drude model, as the conductivity $\sigma \propto \frac{1}{m_{eff}}$ is inversely proportional to the effective masses. This implies that the comparatively higher minimum for the effective masses in the tetragonal phase leads to a corresponding lowered maximum for the conductivity, resulting in a bottleneck for charge carrier transport. Based on this simplified description, the reduced symmetry and resulting band-edge dispersion in monoclinic \ce{BiVO4} imply more favorable charge transport characteristics than in the tetragonal phase. This could provide a plausible microscopic explanation for the experimentally observed differences in their photocatalytic and photoelectrochemical performance characteristics.

It must be noted, although polaronic effects likely play a significant role in \ce{BiVO4}, resulting in a renormalization of the effective masses, this may not impact the qualitative differences between the phases. As the correct description of polarons in \ce{BiVO4} is not yet established \cite{Liu2022}, it is not yet possible to accurately calculate the mobilities for \ce{BiVO4}. Provided the polaronic coupling does not heavily depend on the phase - as appears likely due to their highly similar structures - it appears plausible that the relationships of the effective masses in the pristine structures may be approximately conserved in the renormalized polaron masses. In particular, averaging the conductivity contributions at each symmetry-equivalent extremum likely has a similar effect of suppressing the most efficient transport directions in tetragonal \ce{BiVO4} when transport is governed by polarons, although more advanced models may be required when charge-carriers cannot be assumed to remain close to the band extrema. 

\begin{table}[ht]
\centering
\caption{Band-edge positions in fractional coordinates and effective masses%
\footnote{All effective masses are given in units of the free-electron mass $m_e$.}
in monoclinic and tetragonal \ce{BiVO4}, including symmetry-averaged values%
\footnote{For tetragonal \ce{BiVO4}, the effective-mass values obtained after averaging over all symmetry-equivalent VBM and CBM extrema according to
$\overline{\frac{1}{m^*}}
=
\frac{1}{\sum_g 1}
\sum_g
R_g \frac{1}{m^*} R_g^T$,
where $R_g$ denotes the rotation matrix associated with each unique extremum $g$, are given in parentheses.}.}
\label{tab:band_structure}
\vspace{2mm}

\begin{tabular}{c|c|c}
    \toprule
        & \ce{monoclinic} & \ce{tetragonal} \\
    \midrule
        VBM & $(0.09,\,0.09,\,0.21)$ & $(-0.01,\,0.01,\,0.34)$ \\
        CBM & $(0.18,\,0.17,\,0.13)$ & $(0.20,\,-0.20,\,0.29)$ \\
    \midrule
        $E_{\mathrm{g}}\,(\text{static})$
        & \SI{3.22}{\electronvolt}
        & \SI{3.15}{\electronvolt} \\

        $E_{\mathrm{g}}\,(\SI{300}{\kelvin})$
        & \SI{2.81}{\electronvolt}
        & \SI{2.74}{\electronvolt} \\

        $E_{\mathrm{g}}^{\mathrm{opt}}\,(\SI{300}{\kelvin})$
        & \SI{2.60}{\electronvolt}
        & \SI{2.51}{\electronvolt} \\
    \midrule
        $m^*_{e,1}$ & 3.11 & 4.91 (4.91) \\
        $m^*_{e,2}$ & 1.14 & 1.28 (0.82) \\
        $m^*_{e,3}$ & 0.54 & 0.60 (0.82) \\
    \midrule
        $m^*_{h,1}$ & 4.79 & 3.90 (0.98) \\
        $m^*_{h,2}$ & 1.20 & 1.33 (1.33) \\
        $m^*_{h,3}$ & 0.43 & 0.56 (0.98) \\
    \bottomrule
\end{tabular}
\end{table}

While the effective masses provide a useful basis for evaluating mobilities and gaining initial insights into the dispersion at the band edges, they are limited to the parabolic region close to the extrema. To gain a more general understanding of the shape of the band edges in reciprocal space, we calculated the isosurfaces corresponding to $\pm k_BT$ at room temperature around the band extrema (see Figure \ref{fig:band_edges}). Usually, this would be achieved via band-structure plots, but for complex band edges such as the ones in \ce{BiVO4}, the representation along several 1D paths cannot saliently capture all details, as we show in the SI. Using the isosurfaces near the band edges, for a parabolic dispersion, we expect to see isosurfaces shaped as simple ellipsoids centered on the band extrema, with deviations implying a breakdown of the effective mass approximation. In the monoclinic phase (panel a), the hole isosurfaces are strongly anisotropic, and have pronounced tails deviating slightly from a perfect ellipsoid shape. The electron dispersion near the CBM initially appears to be approximately parabolic. However, conduction band states located at slightly higher energy also appear, and the corresponding isosurface deviates strongly from the ellipsoid expected for a parabolic band, instead resembling a cylinder extending across the $a^*b^*$-plane of $k$-space (see Figure \ref{fig:band_edges}).  In contrast, the tetragonal phase (panel b) only has isosurfaces which closely resemble ellipsoids centered on the band extrema, as expected for a nearly parabolic dispersion. Although they show significant anisotropy, there are no features implying a breakdown of the effective-mass approximation. Overall, it appears like the effective masses are able to describe the properties of excited carriers well in the tetragonal phase, and are a reasonable description of the holes in the monoclinic phase. However, the appearance of essentially degenerate states slightly above the CBM in the monoclinic phase must lead to stronger electron localization than first suggested by the analysis of the effective masses. This highlights the importance of checking the validity of effective mass calculations by inspecting the near-edge dispersion more generally. The effective masses imply superior electron- and hole transport in monoclinic \ce{BiVO4}, but taking into account stronger electron localization due to flat bands in the monoclinic phase suggests that it is largely hole-transport which boosts the photocatalytic and photo-electrochemical  performance of monoclinic BVO. The electron small polarons on the other hand are likely bound even more tightly in the monoclinic phase than in the tetragonal. 

\subsection{Impact of symmetry breaking on optical absorption}

\begin{figure}
    \centering
    \includegraphics[width=\linewidth,trim={1.0cm 0cm 7.0cm 0.5cm}]{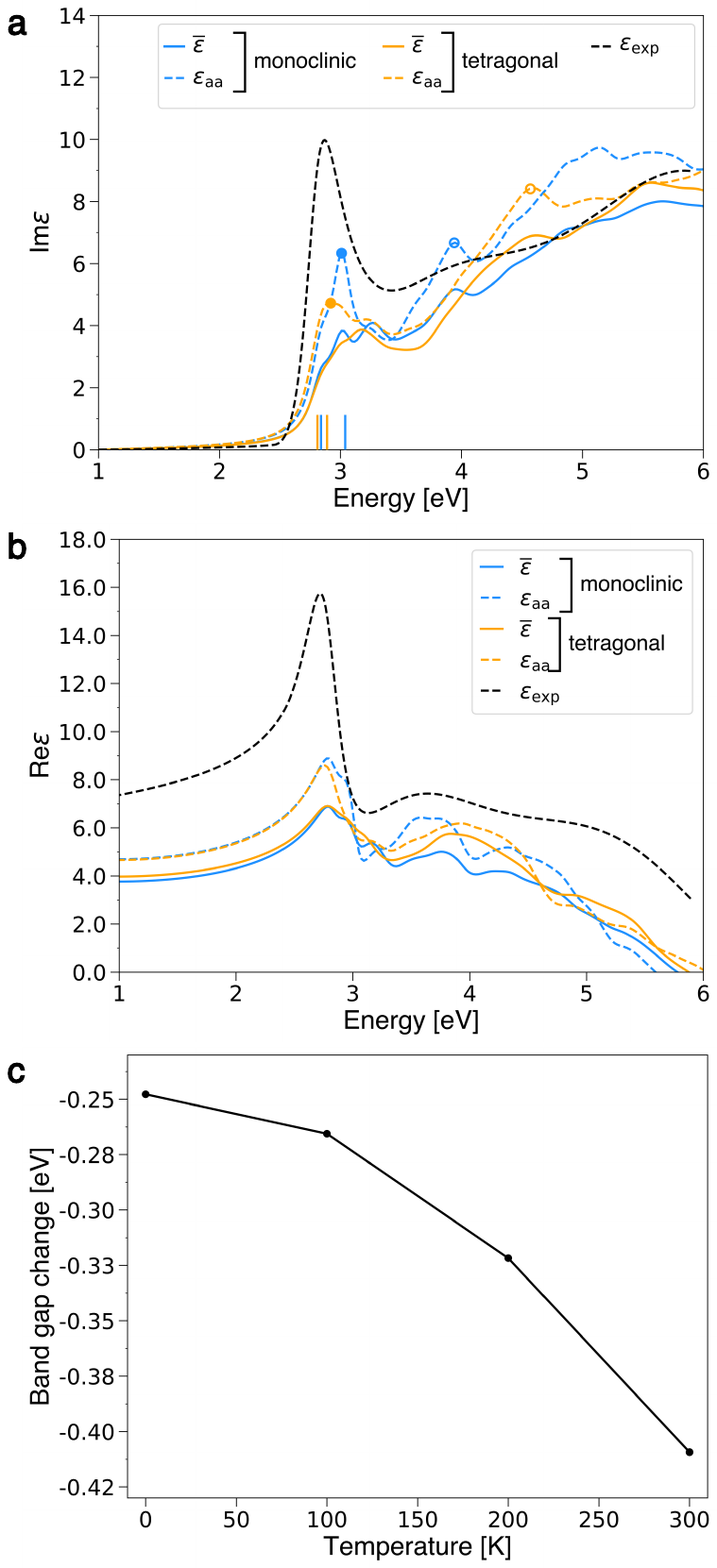}
    \caption{Comparison between the imaginary (a) and real (b) parts of the dielectric function along the lattice vector $\Vec{a}$, $\epsilon_{aa}(\omega)$, as well as the direction-averaged dielectric function $\overline{\epsilon}(\omega)$ for the monoclinic and tetragonal structures of \ce{BiVO4}, as well as the experimental dielectric function for epitaxial monoclinic \ce{BiVO4} reproduced from ref. \cite{Kunzelmann2022}. The blue and orange full circles indicate the energies of the most prominent absorption features. The small vertical blue and orange lines indicate the fine structure of the first absorption peak. (c) Impact of temperature and zero-point fluctuations on the fundamental band gap of monoclinic \ce{BiVO4}.}
    \label{fig:Exciton_and_temperature_change}
\end{figure}

It is well-known that the optical gap of \ce{BiVO4} is strongly affected by both excitonic and thermal effects \cite{Wiktor2017}. Taking into account our findings regarding the impact of the exchange-correlation functional on accurately describing the lattice and electronic structure of \ce{BiVO4}, we now turn to a theoretical investigation of its optical gap. At the HSE06+SOC level without accounting for excitonic and thermal effects, we obtain fundamental band gap values of $3.22$\,eV for monoclinic \ce{BiVO4} and $3.15$\,eV for tetragonal \ce{BiVO4}, as listed in Table \ref{tab:band_structure} for the static minimum energy structure.
To account for excitonic effects in the static minimum energy structure, we calculate the dielectric function of \ce{BiVO4} according to Eq.~\eqref{eq:dielectric function}. From the theoretical absorption spectra $\epsilon_{aa}(\omega)$ shown in Figure \ref{fig:Exciton_and_temperature_change}a, we extract exciton binding energies of $\SI{0.21}{eV}$ for the monoclinic structure and $\SI{0.23}{eV}$ for the tetragonal structure, as discussed in more detail below. These excitonic corrections reduce the HSE06+SOC fundamental gaps to $\SI{3.01}{eV}$ and $\SI{2.92}{eV}$, respectively. This is in good agreement with the estimate of \SI{3.1}{eV} for the optical gap of monoclinic \ce{BiVO4} from \cite{Ohad2023}, showing that HSE06 is comparable to WOT-SRSH in this case. We have even better agreement with the ion-clamped exciton binding energies for monoclinic \ce{BiVO4} of $\SI{0.15}{eV}$ reported in ref. \cite{gant2025}, although we do not account for a further reduction due to lattice dynamics discussed in that work. 

The shape of the imaginary part of the dielectric function is qualitatively similar for monoclinic and tetragonal \ce{BiVO4}, and shows good agreement with the experimental values for the monoclinic phase, as seen in \ref{fig:Exciton_and_temperature_change}a. In both cases, there is significant directional dependence, exemplified by the difference between the dielectric function $\epsilon_{aa}(\omega)$ along the lattice vector $\Vec{a}$, and the directionally averaged $\overline{\epsilon}(\omega)$. The first absorption peak, which was used to determine the optical correction to the fundamental band gap, is composed of two excitonic states. In the monoclinic phase, their binding energies relative to the fundamental gap are $\SI{0.38}{eV}$ and $\SI{0.18}{eV}$, marked by blue vertical lines in \ref{fig:Exciton_and_temperature_change}a. The two states are moved closer together in the tetragonal structure, with values of $\SI{0.34}{eV}$ and $\SI{0.26}{eV}$, marked in orange in \ref{fig:Exciton_and_temperature_change}a. When broadening is added to mimic thermal and finite size effects, they merge into a single peak, marked by the blue and orange circles above the vertical lines.

The most notable difference between spectra for the monoclinic and tetragonal structures appears at higher energies. The second excitonic peak appears at $\SI{3.94}{eV}$ for the monoclinic and at $\SI{4.57}{eV}$ for the tetragonal phase, an increase of $\SI{0.63}{eV}$. The low-frequency response is comparatively unchanged, with only a $\SI{0.02}{eV}$ difference in the binding energy extracted from the first peak. The same is true for the real part of the dielectric function, shown in \ref{fig:Exciton_and_temperature_change}b, demonstrating nearly identical low-frequency dielectric constants. This finding supports the use of the same screening parameter for both phases. However, the strong anisotropy of the dielectric constant seen in \ref{fig:Exciton_and_temperature_change}b does imply that the use of a scalar static dielectric constant to screen the Coulomb interaction \(W\) mentioned in \ref{sec:level1d} may be a source of error in the calculations. Nevertheless, the dielectric constants obtained from the optical response, $4.7$ along the $\Vec{a}$-direction and $3.8$ for the spatial average, are close to the HSE06+SOC predicted value of 5.35. All aforementioned values for the dielectric constant lead to bound excitons, as is discussed in more detail in the SI, further confirming that these results are qualitatively stable. Finally, although the experimental amplitude for the real dielectric function is systematically underestimated, the line shape shows excellent agreement. 

\begin{figure*}[t]
    \centering
    \includegraphics[width=\textwidth]{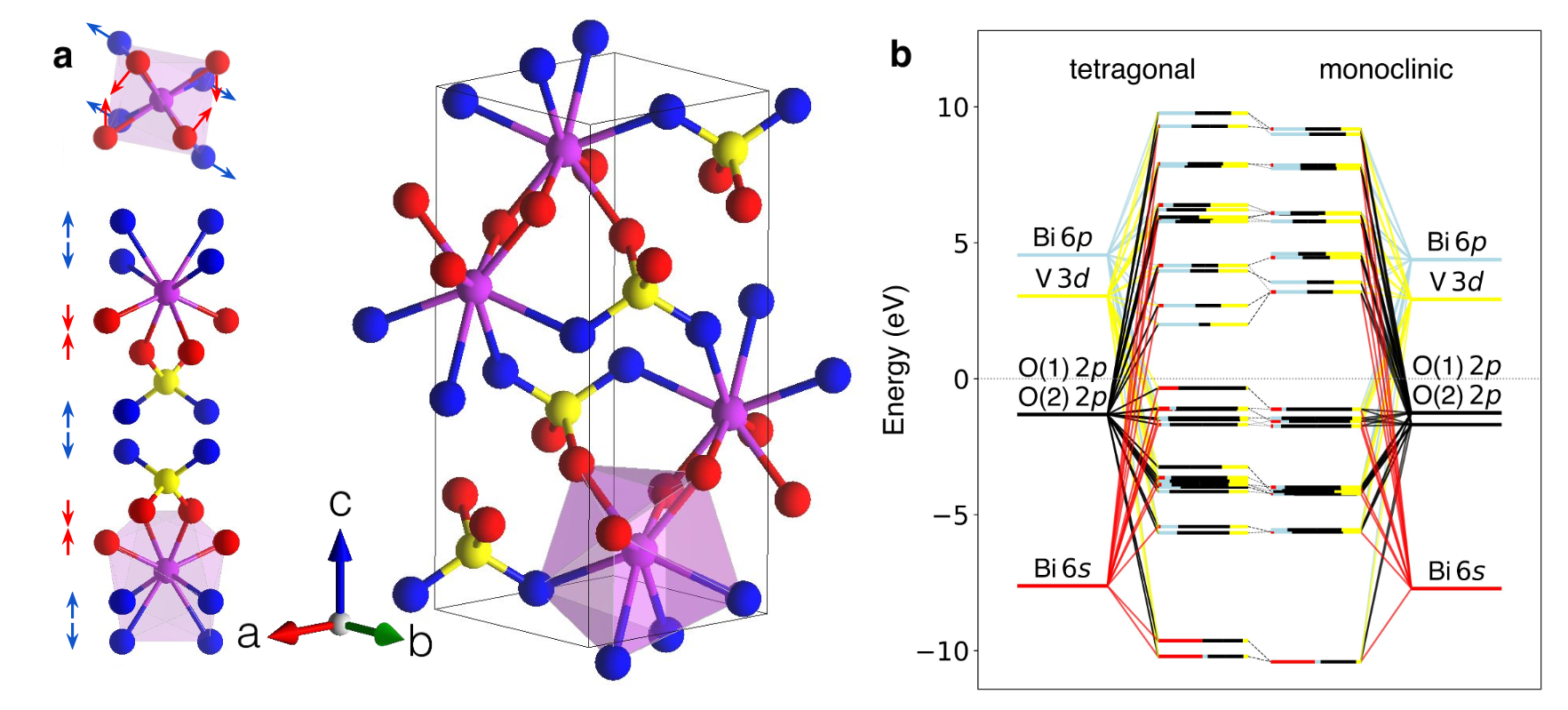}
    \caption{(a) The crystal structure of \ce{BiVO4}, with bismuth (purple), vanadium (yellow), and oxygen (red and blue) atoms. Oxygen atoms that move closer to the Bi atoms in the monoclinic structure (O1) are shown in red, whereas those with increased Bi--O bond lengths (O2) are shown in blue. In the top left corner, the structural change of the \ce{BiO8}-group projected into the $\Vec{a}\Vec{b}$-plane is shown, and the changes when projected into the $\Vec{a}\Vec{c}$-plane are shown below. 
    (b) A simplified bonding diagram illustrating the two mechanisms responsible for stabilization of the monoclinic distortion. The O1 atoms move closer to the Bi atoms, thereby reducing their on-site energies, while symmetry breaking also leads to enhanced Bi6s--Bi6p hybridization.}
    \label{fig:crystal_changes}
\end{figure*}

Having neglected thermal effects so far, we now compute their impact on the fundamental band gap via Monte Carlo simulations. We find that zero-point fluctuations already lead to a reduction of the gap by $\SI{-0.25}{eV}$, as shown in Figure \ref{fig:Exciton_and_temperature_change}c, 
in reasonable agreement with ref.~\cite{Wiktor2017}, which reported a reduction in the fundamental band gap by $\SI{-0.22}{eV}$ due to nuclear quantum effects. However, at $\SI{300}{K}$ we obtain a total reduction of $\SI{0.41}{eV}$, significantly smaller than their value of $\SI{0.92}{eV}$. This difference is likely due to their use of PBE0 for band gap calculations but only PBE for the interatomic forces -- necessitated by the massive computational cost of using (path integral) molecular dynamics -- whereas we consistently used HSE06+SOC throughout, at the cost of neglecting anharmonic effects. 
Both the reduced slope near $\SI{0}{K}$ and the temperature-induced reduction of the fundamental band gap at room temperature by approximately $\SI{0.2}{eV}$ are in excellent agreement with the experimental results found in ref.~\cite{David1983}. Thus, these results provide strong corroborating evidence that harmonic Monte Carlo calculations are sufficient to describe the influence of the atomic dynamics on the band gap at room temperature, provided the electronic structure is treated using HSE06+SOC. 

Finally, including both excitonic and thermal effects, we find a total reduction in the HSE06+SOC gap by approximately $\SI{0.62}{eV}$ at \SI{300}{K}, leading to a predicted optical band gap of $\approx \SI{2.60}{eV}$ for monoclinic \ce{BiVO4} at room temperature, while tetragonal \ce{BiVO4} is predicted to exhibit an optical band gap of $\approx \SI{2.51}{eV}$ under the same conditions. Thus, our calculations indicate a difference in the optical  gap of monoclinic and tetragonal \ce{BiVO4} of 0.09 eV, in very good agreement with the experimentally measured difference of 0.07\,eV \cite{Tokunaga2001}. The predicted optical gap of monoclinic \ce{BiVO4} is also in good agreement with the experimentally reported values, which lie between $\SI{2.4}{eV}$ and $\SI{2.5}{eV}$ \cite{Payne2011,Stoughton2013,Cooper2014,Cooper2015,Kunzelmann2022}. It should be noted that summing the excitonic and thermal corrections neglects the impact of lattice vibrations on the excitons themselves, which would lead to a reduction in binding energies according to ref. \cite{gant2025}.
These results show that excitonic effects reduce the fundamental band gap in both monoclinic and tetragonal \ce{BiVO4} in essentially the same manner, indicating that Coulomb interactions give rise to low‑energy excitons with comparable binding energies in both polymorphs. When thermal effects are additionally included, the calculated optical gaps accurately reproduce the measured ones.

\subsection{\label{sec:level2b}Symmetry breaking mechanism}

\begin{figure*}[ht]
    \centering
    \includegraphics[width=\linewidth]{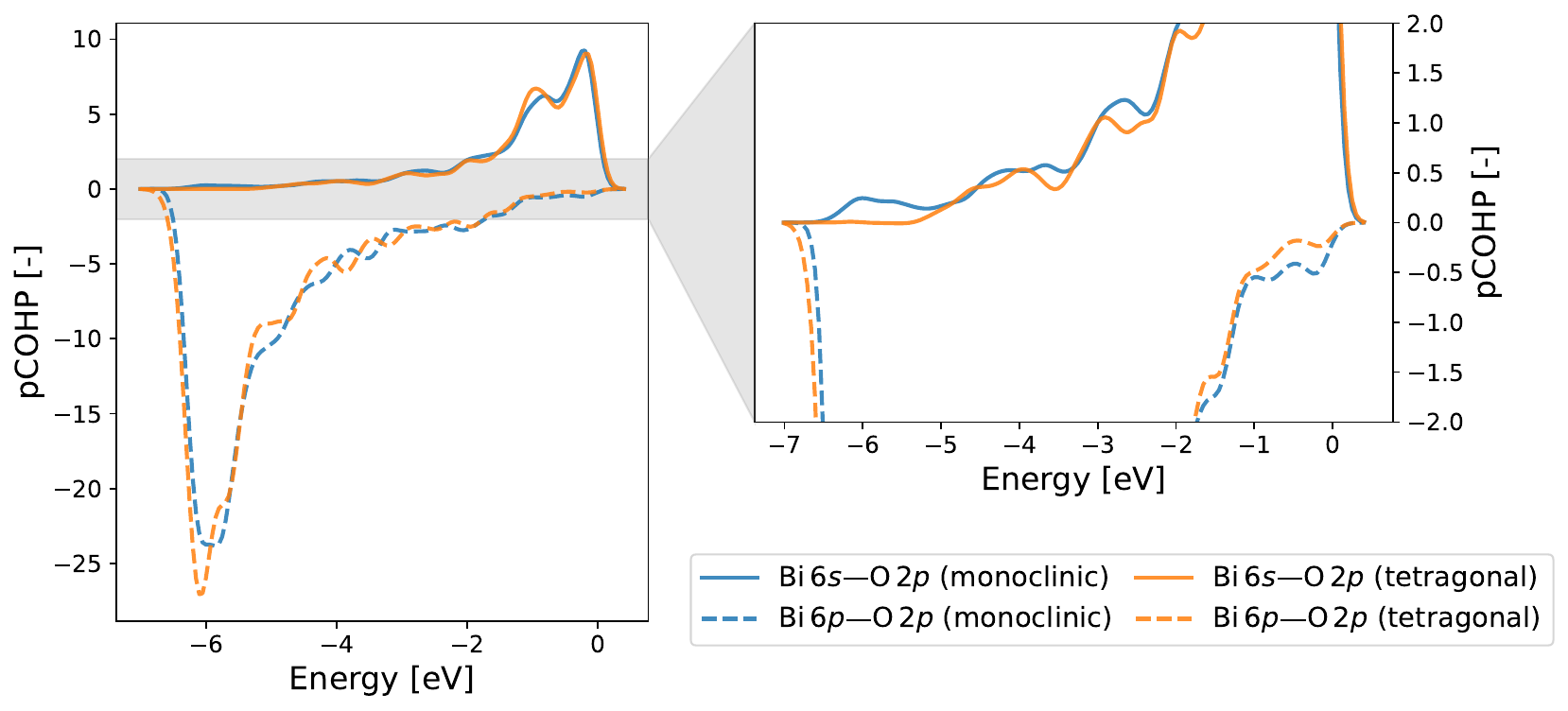}
    \caption{pCOHP calculations for the Bi--O bonds in the monoclinic (blue) and tetragonal (orange) structures of \ce{BiVO4} (the negative values indicate bonding, the positive values antibonding orbital overlap). The left panel indicates the reduction of Bi\,6p--O\,2p bonding contributions at -6 eV and Bi\,6s--O\,2p antibonding contributions near the VBM in the monoclinic phase. The zoomed-in panel on the right highlights the more subtle changes for clarity. }
    \label{fig:pCOHP}
\end{figure*}

Since the monoclinic distortion is associated with improved photocatalytic performance \cite{Kudo99}, it is important to understand the microscopic origin of these functional differences. As a starting point we provide a mechanistic description of the underlying driving forces for the ground state monoclinic distortion itself.
By performing accurate DFT calculations with various hybrid functionals, we identified the approach that most accurately reproduces the ground state structure and corresponding electronic structures. Building on accurate ground-state wave functions at the generalized Kohn-Sham level, we now utilize three tools to identify the key bonding mechanisms responsible for symmetry breaking: the pCOHP formalism, Bader charge analysis and a heuristic description via projection onto a TB Hamiltonian. Finally, we rationalize the necessity of hybrid functionals and SOC by investigating their impact on the uncovered symmetry-breaking mechanisms.

To understand the changes in bonding across the phase transition, we first consider the structural differences between tetragonal and monoclinic \ce{BiVO4} (see Figure \ref{fig:crystal_changes}).
In the tetragonal structure, all oxygen atoms are symmetry-equivalent. The monoclinic distortion primarily changes the Bi--O bonds, with half the Bi--O bonds in the structure shortening and the other half elongating. This results in a layered structure along the $\Vec{c}$-axis in the monoclinic phase, as shown in Figure \ref{fig:crystal_changes}\textbf{a}, where the blue oxygen atoms, labeled as O(2), have a larger distance to the Bi atoms, while the red O atoms, labeled as O(1), move closer to Bi sites.
In contrast, the red O(2) atoms have larger V--O bonds than the blue O(2) atoms.
These structural changes, which primarily occur along the $\Vec{c}$-axis of the conventional cell, are often described as the Bi-atoms moving off-center within the BiO$_8$ dodecahedra. However, projection onto the $ab$-plane reveals that the O(1) atoms also move closer to the Bi atoms along the diagonal between the $\Vec{a}$ and $\Vec{b}$ axes, leading to the increase of the monoclinic angle, $\gamma$, associated with the monoclinic distortion. At the same time, the O(2) atoms move away from the Bi atoms, primarily along the \textbf{a} axis, causing an increase in the lattice constant ratio, $\frac{a}{b}$. Therefore, most structural changes across the phase transition can be understood in terms of shortening of the Bi--O(1) bonds and elongation of the Bi--O(2) bonds. 

Having characterized the structural changes associated with the phase transition, we next analyze their microscopic origin. To this end, we constructed a simplified bonding diagram by transforming the DFT Hamiltonian into an effective TB Hamiltonian (see Methods) for both the tetragonal and monoclinic phases of \ce{BiVO4}, as shown in Figure \ref{fig:crystal_changes}\textbf{b}. As established in ref. \cite{Kweon2012}, covalent bonding between Bi\,6s and O\,2p states leads to bonding states around $\SI{-10}{eV}$, and anti bonding states at the VBM. The bonding states have mostly Bi\,6s character, whereas the antibonding states are dominated by O\,2p contributions. Furthermore, covalent bonding between Bi\,6p, V\,3d and O\,2p leads to occupied bonding states, dominated by O\,2p, between $\SI{-2}{eV}$ and $\SI{-6}{eV}$, with the corresponding antibonding states, dominated by metals, found at and above the CBM.
In the tetragonal structure, hybridization between Bi\,6s and Bi\,6p states is reduced by symmetry. 
In a perfect rotationally symmetric environment, Bi\,6s and 6p states are completely orthogonal. 
However, the centrosymmetric \ce{BiO8}--environment only approximates spherical symmetry, hence, many bands show a small amount of Bi\,6p--Bi\,6s hybridization. The states which do respect the approximate symmetry, seperate into orbitals with mainly Bi6s--O2p and Bi\,6p--O\,2p character. 
Between $\SI{-2}{eV}$ and $\SI{-7}{eV}$ relative to the VBM, bonding Bi\,6p--O\,2p orbitals dominate, while bonding Bi\,6s--O\,2p orbitals are located approximately $\SI{-10}{eV}$ below the VBM. 
At the CBM, V\,3d--Bi\,6p--O\,2p states dominate, with the lowest energy contributions lacking any hybridization with Bi\,6s states.

Breaking the symmetry via the monoclinic distortion enables increased Bi6s--Bi6p hybridization \cite{Walsh2009, Kweon2012}. Mixing of the Bi\,6s and Bi\,6p states introduces additional antibonding Bi\,6s--O\,2p character at the CBM, increasing the energy of these states significantly. The opposite occurs at the VBM, where adding bonding Bi6p-character to the antibonding Bi\,6s--O\,2p states reduces their energy. 
The average energies and weights of the states between $\SI{-2}{eV}$ and $\SI{-7}{eV}$ change only slightly between the two phases, as they are strongly affected by several opposing mechanisms, which will be more thoroughly discussed using the pCOHP and Bader charge-analysis. Finally, the bonding Bi\,6s--O2 p states at $\SI{-10}{eV}$ evolve into more strongly bonding Bi\,6s--Bi\,6p--O\,2p states, with even lower energies.

These trends in the calculated bonding diagrams are corroborated by the pCOHP results shown in Figure \ref{fig:pCOHP}. In monoclinic \ce{BiVO4}, we find increased bonding Bi\,6p--O\,2p contributions above $\SI{-2}{eV}$ relative to the VBM, which is accompanied by a corresponding decrease in antibonding Bi\,6s--O\,2p character in the same energy range. The effect between $\SI{-7}{eV}$ and $\SI{-2}{eV}$ now becomes clear: Increased mixing of the bonding Bi\,6p--O\,2p states with Bi\,6s contributions increases antibonding Bi\,6s--O\,2p, and reduces bonding Bi\,6p--O\,2p contributions. Overall, this demonstrates that symmetry-breaking increases mixing between Bi\,6p- and Bi\,6s states. Since the Bi\,6s-derived antibonding states are mostly concentrated in the range from $\SI{-2}{eV}$ to \SI{0}{eV}, they mix with Bi\,6p states from occupied and unoccupied bands.
Shifting bonding Bi\,6p--O\,2p contributions from unoccupied conduction band states to occupied states lowers the total energy. This reduction in energy is compensated by the shifting of antibonding Bi\,6s--O\,2p contributions from occupied valence band states to unoccupied conduction band states, which then increase in energy. Consistent with this picture, the Bi-O pCOHP integrated up to the Fermi energy, which provides a measure for the overall Bi-O bond energy, is $\SI{6}{meV}$ per bond smaller in monoclinic \ce{BiVO4} than in tetragonal \ce{BiVO4}. This model explains the larger fundamental band gap in monoclinic \ce{BiVO4}, as symmetry breaking lowers the energy of the VBM and increases the energy of the CBM.

\begin{figure*}
    \centering
    \includegraphics[width=\textwidth,trim={3.0cm 3.0cm 3.0cm 3.0cm}]{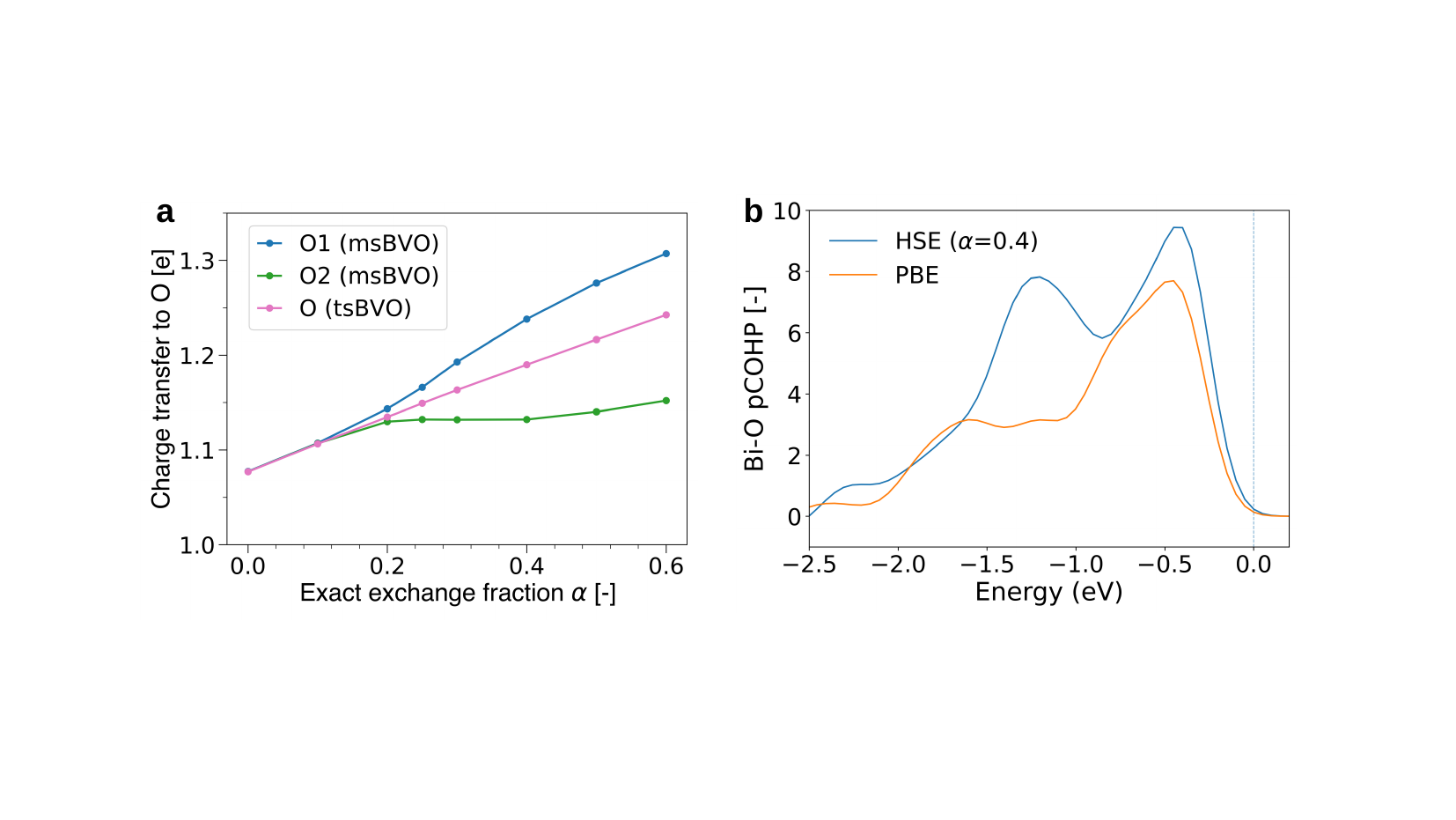}
    \caption{(a) Change in the charge transferred to the oxygen atoms as a function of the exact exchange fraction used in the structural relaxation. (b) Comparison of the near-VBM Bi--O antibonding interaction in the tetragonal phase calculated with PBE and HSE06  with $\alpha = 0.4$, which was chosen to best reproduce experiment while neglecting SOC.}
    \label{fig:Impact_XX_bonding}
\end{figure*}

While the discussion above focused on the covalent bonding changes captured by the molecular orbital diagram and pCOHP analysis, the same structural symmetry breaking also modifies the ionic component of the bonding. The monoclinic distortion renders the oxygen sites inequivalent, as discussed in the structural analysis above, leading to two distinct groups of O atoms with different on-site energy levels. The O(1) atoms, which form the shortest Bi--O bonds, exhibit lower on-site energies because of their close proximity to the positively charged Bi atoms and, therefore, accumulate more charge than the more distant O(2) sites. This picture is supported by Bader charge analysis, which reveals a difference of 0.1 electrons per atom between O(1) and O(2) sites. Hence, the enhanced charge transfer to the proximal oxygen atoms strengthens the ionic bonding in the monoclinic phase compared to the tetragonal phase.

The combined effect of ionic and covalent contributions can be understood consistently within the TB framework used to construct the bonding diagram (see Figure \ref{fig:crystal_changes}\textbf{b}). In this representation, ionic effects primarily modify the on-site energies, whereas covalent bonding arises from hybridization encoded in the off-diagonal elements of the Hamiltonian. As a result, lowering the on-site energies stabilizes both occupied and unoccupied states with contributions from the affected atomic orbitals. In contrast, covalent hybridization necessarily lowers the energies of occupied states while raising those of unoccupied states, reflecting the bonding–antibonding splitting. This also explains the energy of the states between $\SI{-2}{eV}$ and $\SI{-7}{eV}$ remaining constant across both phases -- as the pCOHP analysis shows, additional Bi\,6s--O\,2p antibonding contributions raise the energy in the monoclinic phase, but the high weight of O\,2p states, which are reduced in energy by the overall lowering of on-site energies, offsets this effect. 

In addition to the Bi--O interactions described above, we identify two additional effects that play a non-neglibile role in the symmetry breaking mechanism. First, our pCOHP calculations indicate that the V--O bonds are also strengthened in monoclinic \ce{BiVO4} since the pCOHP integrated up to the Fermi energy increases by $\SI{17}{meV}$ per bond. This can be seen in the bonding diagram in \ref{fig:crystal_changes}b, where the energy of the V\,3d--O\,2p bonding state at $\approx \SI{-3}{eV}$ is significantly lowered, and hybridization with Bi6p increases slightly. This energy-reduction likely has two causes: First, V\,3d--O\,2p bonding is strengthened by reducing the V-O1 bond length, and secondly, hybridization between V\,3d and Bi\,6p is enabled, as the V-atom moves closer to the Bi-atom due to shortened Bi-O1 and V-O1 bonds. Second, we saw in Figure \ref{fig:Monoclinic_Angle} that inclusion of SOC energetically stabilizes the monoclinic phase, but does not by itself induce symmetry breaking. This stabilizing role of SOC can be understood in terms of its effect on orbital hybridization. According to ref. \cite{SOC_impact_bonding2022}, for bonds between half-filled p orbitals, SOC enables a redistribution between bonding and antibonding molecular orbitals, which weakens the overall bonding and allows bond elongation. In \ce{BiVO4}, this effect reduces the energy required for the elongation of the Bi--O(2) bonds, enabling symmetry breaking at lower exact exchange fractions. 

Lastly, we address the open mechanistic question of how and why exact exchange impacts symmetry-breaking in \ce{BiVO4}. The answer is derived from our description of the microscopic bonding interactions underlying the phase transition. First, we show in Figure \ref{fig:Impact_XX_bonding}a that the lowering of on-site energies on the O(1)-atoms in the monoclinic phase, as discussed above, is accompanied by accumulation of charge on these sites. Such charge localization involves a spurious energy penalty in semilocal DFT functionals due to self-interaction errors \footnote{In this work, we use the term 'self-interaction errors' to refer to both the one-electron self-interaction error, as well as the many-electron self-interaction error stemming from deviations from piecewise linearity in semi-local DFT as described in ref. \cite{Kronik2020}}. Hence, increasing the amount of exact exchange in the calculations also increases the driving force for the transition into the monoclinic structure. We extrapolate that semilocal DFT will generally underestimate symmetry breaking in electron-ion systems, especially when the distortion leads to a redistribution of charge relative to the higher symmetry-state. This is in excellent agreement with the dependence of the soft mode on the exchange-correlation functional reported in \ce{SrTiO3}, \ce{KTaO3} and \ce{BaTiO3} found in ref. \cite{Verdi2023}. Additionally, it had been hypothesized in ref. \cite{Kweon2012} that increasing the amount of exact exchange strengthens the Bi--O interaction at the VBM, thereby increasing the impact of Bi6s--Bi6p hybridization in the monoclinic phase. To test this hypothesis, we calculated the Bi--O pCOHP near the VBM for tetragonal \ce{BiVO4} using both PBE and a HSE-like range-separated hybrid with $\alpha = 0.4$, as shown in Figure \ref{fig:Impact_XX_bonding}b. We find that the antibonding interaction at the VBM is substantially stronger when $\alpha = 0.4$, despite the crystal structures being identical. 
While the microscopic origin of this behavior remains unresolved, the present results suggest that hybrid functionals may systematically alter inter-orbital hybridization and bonding interactions in complex oxides, opening an interesting direction for future theoretical work.

\section{\label{sec:level3}Conclusions}

In conclusion, we investigated the interplay between electronic structure, chemical bonding, and lattice symmetry in the prototypical photocatalyst \ce{BiVO4} using first-principles calculations.
We find that the crystal symmetry and electronic structure are intimately linked in \ce{BiVO4} and, as a result, the predicted ground state structure depends sensitively on the level of theory employed. By systematically analyzing hybrid functionals with and without SOC, we identified a theoretical approach capable of accurately reproducing the experimentally observed monoclinic ground state and its associated structural parameters. In addition, we find significant differences between the electronic structures of monoclinic and tetragonal \ce{BiVO4}, including differences in the energies, positions, dispersions, and degeneracies of the VBM and CBM. The symmetry-adjusted effective masses suggest more favorable transport properties in the monoclinic phase, in agreement with the literature. Analysis of the full near-edge fermi-surfaces seems to suggest that despite the favourable effective masses, the monoclinic structure forms more strongly bound electron polarons, implying that experimental observations of superior photocatalytic and photoelectrochemical performance of monoclinic \ce{BiVO4} are largely due to superior hole transport. The optical response of the material is only mildly affected by symmetry breaking at lower energies, despite the marked differences in the shapes and positions of the band edges, though large differences emerge at higher energies. 

Many of the observed differences in the optoelectronic properties of the two phases can be directly connected to the microscopic mechanisms responsible for the monoclinic distortion itself. Enhanced covalent Bi--O bonding leads to an opening of the band gap in monoclinic \ce{BiVO4}, and is dominated by the antibonding Bi--O interaction at the VBM, which requires hybrid functionals and SOC to be described correctly. The size and positioning of the near-edge regions in the conduction bands of monoclinic \ce{BiVO4} support strongly localized states that favor small polaron formation. However, even in the electronic ground state, charge transfer to the oxygen atoms displaced closer to the formerly centro-symmetric Bi sites contributes to stabilization of the monoclinic phase through enhanced ionic bonding. This finding demonstrates a causal link between charge localization and symmetry breaking, explaining the strong dependence of the monoclinic distortion on the amount of exact exchange included in the calculations. 

Self-interaction errors, which appear in semi-local DFT\cite{Kronik2020}, spuriously increase the energies of structures in which symmetry-breaking-induced charge transfer leads to stronger localization of electronic charge. We speculate that this mechanism likely applies to the description of any material in which symmetry breaking generates several inequivalent local atomic environments, and therefore warrants a careful examination of the use of semi-local DFT in studies of lattice dynamics and phase stability. Finally, the exchange-correlation functional that best describes charge localization and structural parameters also yields excellent agreement for the band gap once excitonic and thermal effects are taken into account. This suggests that tuning the fraction of exact exchange to structural observables may provide a more robust strategy than direct tuning to the band gap, and may be more broadly generalizable to other materials of the same class. 

\begin{acknowledgments}

Funding provided by Germany's Excellence Strategy (EXC 2089/2-390776260), as well as from TUM.solar in the context of the Bavarian Collaborative Research Project Solar Technologies Go Hybrid (SolTech) is gratefully acknowledged. The authors thank the Gauss Centre for Supercomputing e.V. for funding this project by providing computing time through the John von Neumann Institute for Computing on the GCS Supercomputer JUWELS at Jülich Supercomputing Centre. We also gratefully acknowledge grants for computer time from the Leibniz Supercomputing Centre in Garching. 

\end{acknowledgments}

    \textbf{Supporting Information}: Convergence tests for plane-wave cutoff and k-grid, comparison of RMSD of bond-lengths from experiment for various fractions of exact exchange, discussion of impact of dielectric constant on optical absorption, band structure diagrams (PDF). 

    \textbf{Data availability:} The data underlying this study are openly available via Zenodo at \url{https://doi.org/10.5281/zenodo.21108667}. 

\bibliographystyle{achemso}
\bibliography{reference}

\end{document}